%% file: paper.tex
\documentclass[sigconf]{acmart}

\usepackage{booktabs}
\usepackage{tabularx}
\usepackage{listings}
\usepackage{hyperref}

\newenvironment{smitemize}%
  {\begin{list}{$\bullet$}{\leftmargin=0em \itemindent=0em}%
     {\setlength{\parsep}{0pt}%
	  \setlength{\topsep}{0pt}%
	  \setlength{\itemsep}{0pt}}}%
  {\end{list}}

\newcommand{\ie}{i.e.,\@~}
\newcommand{\eg}{e.g.,\@~}
\newcommand{\etal}{et al.\@~}

\newcommand{\noindgras}[1]{\noindent{\bf #1}}

\definecolor{codegreen}{rgb}{0,0.6,0}
\definecolor{codegray}{rgb}{0.5,0.5,0.5}
\definecolor{codepurple}{rgb}{0.58,0,0.82}
\definecolor{backcolour}{rgb}{0.95,0.95,0.92}

\lstdefinestyle{mystyle}{
	backgroundcolor=\color{backcolour},
	commentstyle=\color{codegreen},
	keywordstyle=\color{magenta},
	numberstyle=\tiny\color{codegray},
	stringstyle=\color{codepurple},
	basicstyle=\scriptsize\ttfamily,
	breakatwhitespace=false,
	breaklines=true,
	captionpos=b,
	keepspaces=true,
	numbers=left,
	numbersep=5pt,
	showspaces=false,
	showstringspaces=false,
	showtabs=false,
	tabsize=2
}

\newcommand{\xanthus}{\textsc{\fontfamily{qcr}Xanthus}}
\newcommand{\unicorn}{\textsc{\fontfamily{qcr}Unicorn}}

\newcommand{\xmark}{{\color{red}$\times$}}
\newcommand{\vmark}{{\color{codegreen}$\checkmark$}}

\copyrightyear{2020} 
\acmYear{2020} 
\setcopyright{acmcopyright}\acmConference[P-RECS '20]{3rd International Workshop on Practical Reproducible Evaluation of Computer Systems}{June 23, 2020}{Stockholm, Sweden}
\acmBooktitle{3rd International Workshop on Practical Reproducible Evaluation of Computer Systems (P-RECS '20), June 23, 2020, Stockholm, Sweden}
\acmPrice{15.00}
\acmDOI{10.1145/3391800.3398175}
\acmISBN{978-1-4503-7977-9/20/06}

\begin{document}
\fancyhead{}


\title{\xanthus : Push-button Orchestration of Host Provenance Data Collection}

\author{Xueyuan Han}
\affiliation{%
  \institution{Harvard University}
  \city{Cambridge, MA}
  \country{USA}}
\email{hanx@g.harvard.edu}

\author{James Mickens}
\affiliation{%
  \institution{Harvard University}
  \city{Cambridge, MA}
  \country{USA}}
\email{mickens@g.harvard.edu}

\author{Ashish Gehani}
\affiliation{%
  \institution{SRI International}
  \city{Menlo Park, CA}
  \country{USA}}
\email{ashish.gehani@sri.com}

\author{Margo Seltzer}
\affiliation{%
  \institution{University of British Columbia}
  \city{Vancouver, BC}
  \country{Canada}}
\email{mseltzer@cs.ubc.ca}

\author{Thomas Pasquier}
\affiliation{%
  \institution{University of Bristol}
  \city{Bristol}
  \country{UK}}
\email{thomas.pasquier@bristol.ac.uk}

\renewcommand{\shortauthors}{X. Han et al.}

\begin{abstract}
\input{abstract}
\end{abstract}



\maketitle

\section{Introduction}
\label{sec:introduction}
\input{introduction}

\section{Motivation}
\label{sec:motivation}
\input{motivation}

\section{\xanthus~Framework}
\label{sec:xanthus}
\input{framework}

\section{Related Work}
\label{sec:rw}
\input{rw}

\vspace{-15pt}

\section{Discussion}
\label{sec:discussion}
\input{discussion}

\section{Conclusion}
\label{sec:conclusion}
\input{conclusion}

\section*{Availability}
\input{availability}

\begin{acks}
\input{acknowledgements}
\end{acks}

\bibliographystyle{plain}
\bibliography{biblio}

\end{document}

%% file: abstract.tex
Host-based anomaly detectors
generate alarms by
inspecting audit logs for suspicious behavior.
Unfortunately,
evaluating these anomaly detectors is hard.
There are few high-quality, publicly-available audit logs,
and there are no pre-existing frameworks
that enable push-button creation
of realistic system traces.
To make trace generation easier,
we created \xanthus,
an automated tool
that orchestrates virtual machines
to generate realistic audit logs.
Using \xanthus' simple management interface,
administrators select a base VM image,
configure a particular tracing framework to use within that VM,
and define post-launch scripts
that collect and save trace data.
Once data collection is finished,
\xanthus~creates a self-describing archive,
which contains the VM,
its configuration parameters,
and the collected trace data.
We demonstrate that \xanthus~hides
many of the tedious (yet subtle) orchestration tasks
that humans often get wrong;
\xanthus~avoids mistakes that
lead to non-replicable experiments.

%% file: introduction.tex
Host-based intrusion detectors
sift through audit data for signs of attack.
Training and evaluating such detectors
requires trace data.
Unfortunately,
the security community suffers from a lack of
publicly-available, high-quality datasets~\cite{tavallaee2010toward}.
For example,
DARPA's IDEVAL traces~\cite{lippmann20001999}
are publicly available
but suffer from well-known deficiencies
that hurt the realism of the traces~\cite{mchugh2000testing, mahoney2003analysis, maggi2010detecting}.
However,
academics continue to use these traces~\cite{elshafie2019improving, illy2019securing}
(which are over 20 years old!)
due to a lack of alternative public datasets.

The rise of provenance-based intrusion detection~\cite{han2017frappuccino,han2020unicorn,
milajerdi2019holmes,wang2020you,hassan2018towards, hassan2019nodoze}
has emphasized the dearth of realistic, openly-available traces.
Data provenance~\cite{han2018provenance} is
a particular type of audit data that
uses a graph to describe
the interaction histories of host objects such as
files, processes, and network connections.
The typical workflow to evaluate such detection systems consists of three steps:
1) trace benign and attack workloads to construct a training dataset,
2) build a model based on the training traces, and
3) trace new scenarios on which to test the model.
In theory, publicly released datasets are the output of the first step.

While evaluating \unicorn,
our own provenance-based intrusion detection system (IDS)~\cite{han2020unicorn},
we repeatedly found that released traces were insufficient for our purposes.
For example, DARPA's Transparent Computing dataset~\cite{darpatc} contains only attack scenario traces;
the StreamSpot dataset~\cite{manzoor2016fast} was pre-processed, removing key information.
These are all symptoms of a fundamental problem: each IDS typically
requires a specific kind of trace data, and
published traces are specific to the system for which they were originally designed.
Unlike conventional machine learning applications, where training data consists of a
set of samples and their labels, the ``samples'' in this case are large, complex, non-standard
traces.
To address this mismatch we facilitate faithful replication of both the training and
test data.
In other words, \xanthus~enables \emph{replicability}
~\cite{national2019reproducibility} of both training and test workloads for the
evaluation of provenance-based IDSes.

\xanthus~is a framework for collecting host-based provenance datasets, which
automates:
(1) configuring a data collection framework,
(2) recording data using that framework, and
(3) publishing the results. 
During the configuration stage,
\xanthus~creates VMs with
a deterministic set of initial states defined by
user-provided scripts and
a specific provenance tracking framework
(\eg SPADE~\cite{gehani2012spade} or CamFlow~\cite{pasquier2017practical,pasquier2018ccs}).
\xanthus~saves these images for repeated use.
Then, in the recording phase,
\xanthus~runs a specified workload, which can include hooks for additional scripts
that control the monitoring infrastructure in real time.
When execution completes,
\xanthus~bundles the data, the \xanthus~scripts,
and the \xanthus~configuration files
into a single archive
and publishes the archive (\eg on a configured data repository such as \texttt{Dataverse} or on \texttt{GitHub}).
Other researchers can download the archive
to validate correctness of the collected traces,
replay the workloads with different auditing systems and experimental settings
(\eg with or without attacks),
or replay the saved traces to an analysis tool.
For large-scale experiments,
it works seamlessly with Amazon Elastic Compute Cloud (EC2).


%% file: motivation.tex
Intrusion detection introduces a number of challenges not encountered in
other replicability scenarios.
Provenance systems interoperate in specific ways with the host operating system, and each attack
scenario relies on operating system, library, and application versions.
We use our experience trying to evaluate \unicorn~using public datasets
to motivate \xanthus' key design features.

\subsection{Provenance-Based Intrusion Detection}
\label{sec:motivation:provenance}
Provenance-based IDSes~\cite{han2020unicorn, wang2020you, milajerdi2019holmes} often perform
graph analysis on provenance graphs,
in which vertices represent processes, users,
or kernel resources (\eg inodes) and
edges represent system-call-induced interactions.
However, depending on factors such as security end goals, analysis scope, and runtime performance concerns,
detection systems adopt different capture mechanisms and assume distinct graph semantics;
those that use the same capture infrastructure might focus on different subsets of data that they deem relevant to their analysis.
Meanwhile, security researchers are still developing
new graph models~\cite{darpatc} describing abstract execution semantics,
hoping to facilitate future analysis with better system visibility.
Consequently,
effectively sharing data and enabling data reuse becomes challenging, which is why
\xanthus~facilitates replication of the workload that generates the data instead.

\noindgras{\unicorn.}
\unicorn~\cite{han2020unicorn} is a host IDS that
uses provenance graphs as input.
It leverages the state-of-the-art
system-level provenance tracing frameworks~\cite{pasquier2017practical, gehani2012spade, pohly2012hi}
to model
data flows across an entire system
via kernel resources such as inodes and sockets.
These frameworks
not only interpose on system call invocations,
but also understand the \textit{semantics} of system calls.
For example,
CamFlow~\cite{pasquier2017practical}
can trace how the contents of an incoming network packet
flow into a process via \texttt{recv()}
and out of the process
via a subsequent \texttt{write()} to a disk file.
\unicorn~summarizes benign system execution through
efficient graph compression to model normal host behavior
and defines a similarity metric that quantifies
the deviation of the host's current execution
from its model.
It detects anomalies when the system behaves significantly
differently from its norm.
We use \unicorn~as an example throughout~\autoref{sec:xanthus}.

\captionsetup{aboveskip=-5pt}
\begin{table}
\tiny
\centering
\begin{tabularx}{\linewidth}{lcccccc} \toprule
    \vtop{\hbox{\strut Provenance}\hbox{\strut Tool}} & \vtop{\hbox{\strut Open}\hbox{\strut Source}} & \vtop{\hbox{\strut Actively}\hbox{\strut Maintained}} & \vtop{\hbox{\strut Tagged}\hbox{\strut Release}} & \vtop{\hbox{\strut Binary}\hbox{\strut Release}} & \vtop{\hbox{\strut Documentation}\hbox{\strut Support}} & \vtop{\hbox{\strut \xanthus}\hbox{\strut Supported}} \\ \midrule
    PASS~\cite{muniswamy2006provenance}  & \xmark & \xmark & \xmark & \xmark & \xmark & \xmark \\
    Story Book~\cite{spillane2009story}  & \xmark   & \xmark & \xmark & \xmark & \xmark & \xmark \\
    Burrito~\cite{guo2012burrito}  & \vmark  & \xmark & \xmark  & \xmark & \xmark  & \xmark \\
    Hi-Fi~\cite{pohly2012hi}  & \vmark & \xmark & \vmark  & \xmark & \vmark & \xmark \\
    LPM~\cite{bates2015trustworthy} & \vmark  & \xmark & \xmark  & \xmark & \vmark & \xmark \\ \midrule
    SPADE~\cite{gehani2012spade}  & \vmark  & \vmark & \vmark  & \xmark & \vmark & \vmark \\
    PVM~\cite{balakrishnan2013opus} (w/ DTrace~\cite{gregg2011dtrace})  & \xmark  & \vmark & \xmark  & \xmark & \xmark  & \vmark \\
    CamFlow~\cite{pasquier2017practical}  & \vmark  & \vmark & \vmark  & \vmark & \vmark & \vmark \\ \bottomrule
\end{tabularx}
\caption{The provenance tools we examined and selected (below the middle bar) that are supported by \xanthus} \label{tab:motivation:provenance}
\end{table}

\subsection{404: Data Not Found}
\label{sec:motivation:404}

Provenance-based intrusion detection
has been studied for a decade.
However,
we were surprised by the scarcity
of publicly-available provenance traces;
even compiling and running
prior tracing frameworks
was challenging.

\subsubsection{DARPA's Transparent Computing Dataset}
\label{sec:motivation:404:darpa}

DARPA's Transparent Computing program
sponsors a wide variety of provenance research.
A primary goal
is to use provenance to detect and analyze advanced persistent threats (APT),
\ie attacks that spread their activity
across a long period of time,
hiding malicious behavior
amid normal system events.
DARPA conducted simulated attacks
on realistic servers
to generate public datasets for researchers.
For example,
in 2018,
DARPA ran a two-week-long engagement~\cite{darpatc}
in which red teams
launched APT attacks on
victim machines running five different provenance tracking frameworks.
Prior to the engagement,
DARPA deployed scripts
to generate innocuous background activity
(\eg simulated user logins to ssh daemons).
Although the collected provenance traces are publicly accessible,
DARPA did not release
the data captured from the innocuous background activity,
which makes it difficult to evaluate anomaly-based intrusion detectors,
since anomalies are defined relative to normal behavior.
We petitioned DARPA for 
details about the background activity
but were unable to obtain
the scripts that generated the activity.

\subsubsection{StreamSpot's Dataset}
\label{sec:motivation:404:streamspot}

StreamSpot~\cite{manzoor2016fast}
is an academic project
that introduced a fast streaming analysis on provenance graphs.
The authors made their evaluation dataset public.
However,
like the DARPA dataset,
it lacks a description of non-anomalous behavior.
The dataset is also pre-processed:
it contains only the provenance information
useful to StreamSpot's algorithm.
Hiding raw trace information
diminishes the value of a dataset,
since different analytics systems might examine different kinds of provenance states.

\subsubsection{Other Datasets}
\label{sec:motivation:404:other}

We surveyed other academic frameworks
for tracking and analyzing provenance~\cite{jiang2006provenance, shu2017long, liu2018towards, hassan2018towards, hassan2019nodoze},
but none were accompanied with public datasets.
The associated papers
did make sincere attempts
to describe attack scenarios evaluated by the authors;
however,
our attempts to replicate even well-described attacks
were time-consuming, labor-intensive,
and often ended in failure.
For example,
Jiang \etal~\cite{jiang2006provenance}
used a virtualization environment called vGround~\cite{jiang2005virtual}
to isolate worms in a realistic but confined environment.
Unfortunately,
neither vGround nor the experimental setup scripts
are publicly available.

\subsection{An Ideal Framework}
\label{sec:motivation:desiderata}

We struggled to locate a high-quality, public dataset to evaluate \unicorn,
while our subsequent manual efforts to create our own datasets were equally frustrating.
Often times, we were unable to repeat the same experiment
using a different tracing framework
due to, \eg unexpected environmental changes,
missing packages that existed in prior runs,
or even lost references to the experiment due to our own carelessness.
Based on our experience,
we designed \xanthus~with the following properties in mind:

\begin{smitemize}
	\item \textbf{Replicability}: The framework must collect
	enough information to allow a third party
	to recreate an experiment 
	so that different graph semantic models can be adopted to describe 
	identical system execution (~\autoref{sec:motivation:provenance}).
	For example,
	it must capture the discrete events or generative models
	associated with both malicious behavior and innocuous background activity.
	It also needs to capture environmental features such as version information for the
	operating system and user-level binaries that were running during an experiment.
	
	\item \textbf{Flexibility}: The framework
	should not make assumptions
	about the downstream data consumers.
	When possible,
	it should emit raw, unprocessed data.
	Storage is cheap; thus,
	it should err on the side
	of collecting too much data, not too little.
	\item \textbf{Longevity}: The framework must collect and publish data
	in a way that is not dependent
	on a particular hosting server or distribution technology.
	An ideal dataset is self-hosting
	in the sense that,
	once a researcher has downloaded the bytes in the dataset,
	minimal additional infrastructure
	should be necessary to analyze the data
	or recreate the experiments that generated the data.
	\item \textbf{Usability}: The framework should provide explicit interfaces
	that allow easy scripting
	to generate host behavior,
	collect trace data, and so on.
	To the greatest extent possible,
	configuring the software
	inside the system to trace
	should be automated.
	Creating a self-hosting archive
	should also be automated.
	\item \textbf{Shareability}: Researchers should be able to exchange entire experimental environments.
	Shareability is enabled by flexibility, longevity, and usability.
\end{smitemize}

%% file: framework.tex
\xanthus~assumes that the downstream analytics system requires 
as input host audit data,
but it is agnostic to the specific tracing system used.
Currently, we focus on capturing system-level provenance data (for \unicorn). 
\autoref{tab:motivation:provenance} outlines a set of criteria
we used to compare and select provenance tools supported by {\xanthus}.

\begin{figure}[h]
\centering
\includegraphics[width=0.9\columnwidth]{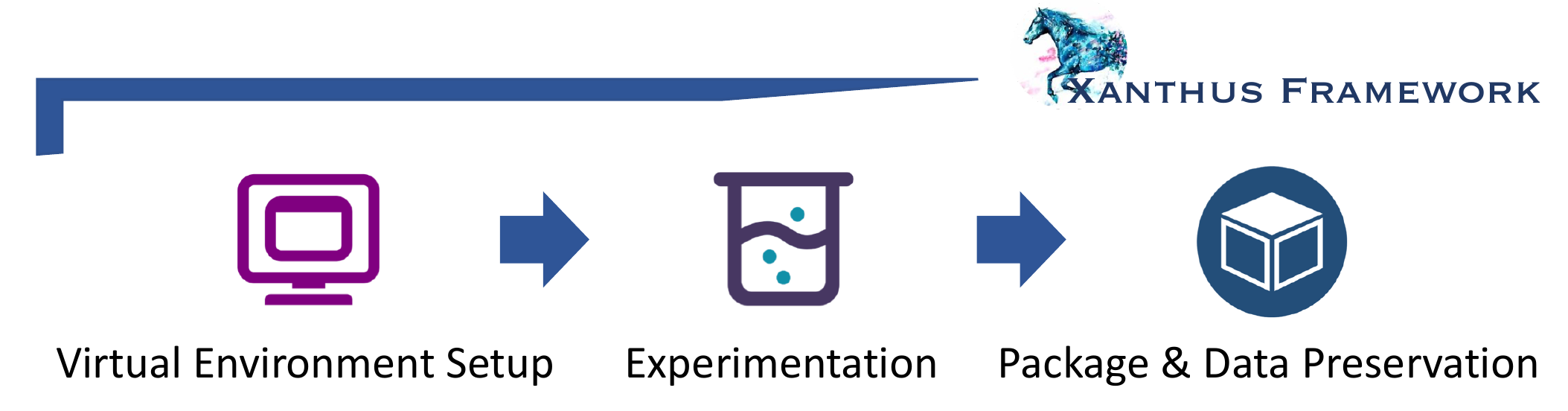}
\caption{\xanthus~framework}
\label{fig:framework}
\end{figure}

\xanthus~is written in \texttt{Ruby} and can be easily installed through \texttt{Ruby}'s package manager \texttt{RubyGems}.
\autoref{fig:framework} shows the three high-level stages that comprise \xanthus.
In the remainder of this section,
we elaborate on each stage with simple code snippets to demonstrate concepts and design decisions.

\vspace{-10pt}

\subsection{Virtual Environment Setup}
\label{sec:framework:setup}
\input{setup}

\subsection{Specifying an Experiment}
\label{sec:framework:experimentation}
\input{experimentation}

\subsection{Package and Data Preservation}
\label{sec:framework:package}
\input{package}

%% file: setup.tex
It is tempting to believe that the script that executes an experiment is
a long-lived artifact.
While a script may provide detailed specifications about a particular
environment, software versions used, and instructions that
automate the experimental setup, 
its correct execution depends
on the availability of those artifacts.
If some version becomes unavailable, replication becomes impossible.

One solution is to provide virtual machine (VM) images encapsulating the correct
environment and software dependencies.
Those materialized images enable immediate replication of an identical working environment.

\begin{lstlisting}[float=h,language=ruby, style=mystyle, caption={Set up a user-configured VM in \xanthus}, label=listing:framework:setup, aboveskip=-5pt, belowskip=-5pt]
config.vm :server do |vm|
    vm.box = 'ubuntu/trusty64'
    vm.ip = '192.168.33.3'
    vm.boxing = [:box_config]
end
config.script :box_config do
    %q{%{
      sudo apt-get install vim apache2
      gem install json rgl mqtt rake bundler
    }}
end
\end{lstlisting}

\xanthus~leverages \texttt{Vagrant} to manage VMs.
Before running an experiment, it creates the necessary VM image(s),
which can be stored locally or shared on VagrantCloud~\cite{vagrantcloud},
an online box repository where users share public boxes.
\xanthus~also supports pre-existing images hosted on VagrantCloud,
which can be further customized through scripts.
For example, in \autoref{listing:framework:setup},
we use the \texttt{ubuntu/trusty64}
image and customize it with the \texttt{box\_config} script.
\texttt{vm.ip} defines the virtual
IP address used during the experiment.
\xanthus~boxes the VM once in the first run and uses the materialized VM afterwards;
this also provides a more
efficient out-of-box experience for those wishing to use the artifact since they
do not have to configure the machine for each experiment.
Users can upload the resulting Vagrant box to VagrantCloud.

\begin{lstlisting}[float=h,language=ruby, style=mystyle, caption={Set up VMs on AWS}, label=listing:framework:aws, aboveskip=-5pt, belowskip=-5pt]
config.vm :aws do |vm|
	vm.on_aws = true
	vm.aws_env_key_id = 'AWS_ACCESS_KEY_ID'
	vm.aws_env_key_secret = 'AWS_SECRET'
	vm.aws_key_pair_name = 'AWS_KEY_PAIR'
	vm.aws_region = 'eu-west-2'
	vm.aws_ami = 'ami-xxxxxxxxxxxxx'
	vm.aws_instance_type = 'i3.large'
	vm.aws_security_group = 'AWS_SG_1'
end
\end{lstlisting}

To enable large-scale, multi-host experiments,
\xanthus~works seamlessly on Amazon EC2 (\autoref{listing:framework:aws});
users simply need to switch to the AWS mode (Line 2) and provide their EC2 credentials in the configuration file
to set up VMs in the cloud.

%% file: experimentation.tex
Each experiment is called a \emph{job},
which consists of instantiation(s) of VM image(s),
execution of user-defined tasks assigned to particular instances,
and management of outputs (\eg to retrieve audit logs).
A \xanthus~workflow is composed of one or more jobs that can be executed multiple times.

\begin{lstlisting}[float=h, language=ruby, style=mystyle, caption={Configure a job in \xanthus}, label=listing:framework:experimentation:job, aboveskip=-5pt, belowskip=-5pt]
config.job :attack do |job|
    job.iterations = 2
    job.tasks = {server: [:server], client: [:pre, :camflow_start, :attack, :camflow_stop, :post]} # VM tasks
    job.inputs = {server: ['ipscan_3.5.5_amd64.deb', 'exploit.py']} # VM inputs
    job.outputs = {client: {config: '/etc/camflow.ini', trace: '/tmp/audit.log'}} # VM outputs
end
\end{lstlisting}

\autoref{listing:framework:experimentation:job} is an example of a job configuration,
in which a job called \texttt{attack} is configured to run twice.
During each iteration,
two VMs, \texttt{server} and \texttt{client}, are instantiated and run their respective task(s).
In Line 3,
\texttt{server} runs a single task, defined in \texttt{config.script :server} (similar syntax as in \autoref{listing:framework:experimentation:attackjob}),
while \texttt{client} has multiple tasks that run sequentially.
A \xanthus~\emph{task} allows users to logically encapsulate a single step in the experiment.
Line 4 defines two inputs to \texttt{server},
a Debian package and a Python script,
while Line 5 shows that we expect two outputs from \texttt{client},
a configuration file and trace data.

\xanthus~is a framework for cybersecurity experiments,
so it is important to ensure easy integration with popular security tools.
We show how users can readily use \xanthus~to retrieve traces during penetration testing using
Metasploit, Armitage, and Cortana.
Without \xanthus, a researcher would have to manually
1) set up an attacker and a victim machine,
2) log onto the attacker machine to configure Metasploit and Armitage,
3) log onto the victim machine to configure its audit system,
4) execute the attack, 
and 5) extract audit data from the victim machine.

\subsubsection*{Security Example}

Metasploit~\cite{kennedy2011metasploit} is a well-known penetration testing framework that helps
security experts verify vulnerabilities, manage security assessments,
and improve security awareness.
Armitage~\cite{kennedy2011metasploit} is a scriptable cyberattack management tool for Metasploit and
a force multiplier (\ie creates synergy) for red team operations.
Cortana is the scripting language behind Armitage that automates the Metasploit framework and creates long running bots.

\begin{lstlisting}[float=h, language=ruby, style=mystyle, caption={Configure attacker and victim VMs}, label=listing:framework:experimentation:vms, aboveskip=-5pt, belowskip=-5pt]
config.vm :victim do |vm|
    vm.box = 'e314c/Metasploitable2'
    vm.version = '0.0.1'
    vm.ip = '192.168.33.8'
end
config.vm :attacker do |vm|
    vm.box = 'Sliim/kali-2018.1-amd64'
    vm.version = '1'
    vm.ip = '192.168.33.10'
end
\end{lstlisting}

In \autoref{listing:framework:experimentation:vms},
we configure an intentionally vulnerable version of Ubuntu Linux VM called Metasploitable~\cite{moore2012metasploitable}.
The Metasploitable VM is designed specifically for testing security tools and demonstrating vulnerabilities.
We then configure a Kali Linux machine, a security-oriented Linux distribution that pre-installs many useful penetration testing tools,
including Metasploit and Armitage.
As we are using existing images from the VagrantCloud, setting them up is trivial,
as illustrated in the listing.

\begin{lstlisting}[float=h, language=ruby, style=mystyle, caption={Configure the adversarial scenario}, label=listing:framework:experimentation:attackjob, aboveskip=-5pt, belowskip=-5pt]
config.job :cortana do |job|
    job.iterations = 1
    job.tasks = {victim: [:actions], attacker: [:attack]}
    job.inputs = {attacker: ['local.prop', 'demo.cna']}
end
config.script :attack do
    %q{%{
      teamserver 192.168.33.10 password &
      java -jar cortana.jar local.prop demo.cna
    }}
end
\end{lstlisting}

Now, let's assume that we wish to simulate an adversarial scenario where the attacker exploits the FTP vulnerability in Metasploitable
and uses the \texttt{vsftpd\_234\_backdoor} module in Metasploit to install a backdoor and create a reverse shell payload to remotely control Metasploitable.
\autoref{listing:framework:experimentation:attackjob} describes the experiment in \xanthus.
The attacker consists of a single task, \texttt{attack}
that launches the attack with a Cortana script.
To run Cortana as a stand-alone script,
the attacker needs to set up an Armitage \texttt{teamserver} locally on the VM.
The user then specifies properties of the team server by placing them in the file
\texttt{local.prop}.
The file \texttt{demo.cna} is the Cortana script that runs the attack (\autoref{listing:framework:experimentation:cortana}).
It creates a virtual Metasploit console that prepares the exploit and configures the payload
(\eg setting up the remote host IP address through \texttt{RHOST}).
To show that our attack succeeds,
the script registers two listeners,
one for when a reverse shell session is open and one for when the shell responds to the \texttt{whoami} command.
When the \texttt{session\_open} event triggers the listener,
the attacker automatically sends a \texttt{whoami} command to the victim and prints victim's response on its console.

\begin{lstlisting}[float=h,language=bash, style=mystyle, caption={The Cortana script}, label=listing:framework:experimentation:cortana, aboveskip=-5pt, belowskip=-5pt]
on shell_whoami {
	println("[ $+ $1 $+ ] I am: $3");
}
on session_open {
	if (!-isshell $1)
		return;
	s_cmd($1, "whoami");
}
$console = console();
cmd($console,"use exploit/unix/ftp/vsftpd_234_backdoor");
cmd_set($console, %(RHOST => "192.168.33.8", RPORT => "21", TARGET => "0",PAYLOAD => "cmd/unix/interact"));
cmd($console, "exploit -j");
\end{lstlisting}

\xanthus~allows
a researcher to easily run similar experiments multiple times with different capture mechanisms
and share precise configurations with others.
\xanthus' modularized design allows researchers
to reuse their experimental setup, simply changing \eg Metasploit's exploit module
to create new experiments.

%% file: package.tex
\xanthus~enables push-button execution of the framework.
The artifacts of the workflow, including user-supplied scripts and packages (as defined in \texttt{job.inputs})
and experimental results and datasets (as defined in \texttt{job.outputs}),
are all bundled and archived locally.

\begin{lstlisting}[float=h,language=ruby, style=mystyle, caption={Configure GitHub and Dataverse}, label=listing:framework:package, aboveskip=-5pt, belowskip=-5pt]
config.github do |github|
  github.repo = '<ADD GITHUB REPO>'
  github.token = '<ADD GITHUB TOKEN>'
end
config.dataverse do |dataverse|
  dataverse.server = '<ADD DATAVERSE BASE URL>'
  dataverse.repo = '<ADD DATAVERSE NAME>'
  dataverse.token = '<ADD DATAVERSE TOKEN>'
  dataverse.subject = '<ADD DATAVERSE SUBJECT>'
end
\end{lstlisting}

\xanthus~allows users to automatically share the collected experimental data.
For example, 
if the user provides a GitHub repository address and an access token,
it pushes the archive to GitHub automatically
using Git Large File Storage (\autoref{listing:framework:package}).
\xanthus~also supports automatic sharing via Dataverse~\cite{king2007introduction},
and we are working on providing more archiving options.
We have made an example archive available at \url{https://github.com/margoseltzer/wget-apt}.
The archive contains a \texttt{.xanthus} file for push-button replicability.
The \texttt{.xanthus} file is the central orchestration file
that controls the entire pipeline described in this section.
It contains metadata describing the experiments and 
actionable instructions to 1) generate VM images,
2) schedule tasks,
3) setup experiments, and
4) store and upload data.

%% file: rw.tex
We are not the first to observe that
cybersecurity research is threatened
by a lack of high-quality, easily-accessible datasets.
For example,
Ghorbani \etal~\cite{sharafaldin2018toward} evaluated
11 publicly-available traces
used by intrusion detection researchers and
concluded that
none of the traces were comprehensive or reliable.
Ghorbani \etal introduced their own dataset (CICIDS2017)
that leveraged their prior work
on systematic generation of IDS traces~\cite{shiravi2012toward}.
However,
the collection of the CICIDS2017 trace
was manually orchestrated (and thus non-replicable).

Despite the power of host-based intrusion detection,
the security community
has traditionally paid more attention to
network traces than host ones.
This bias may reflect
the fact that host-based IDSes
are more recent inventions.
DARPA IDEVAL is a well-known host trace,
but it has various deficiencies,
such as poor diversity of executed programs~\cite{maggi2010detecting}.
We know of only one more host-based dataset
that is widely used---the University of New Mexico dataset~\cite{unm}.
However,
this dataset suffers from similar problems
that hurt the realism of the trace~\cite{pendleton2017dataset}.
Other publicly-available host traces
are either application-specific~\cite{murtaza2013host}
or suffer from low attack diversity and coarse-grained trace information~\cite{creech2013generation};
these datasets are studied
by only a few papers~\cite{haider2017generating, haider2016windows}.
Due to the lack of high-quality datasets,
many evaluations of host-based IDSes
use private datasets~\cite{lichodzijewski2002host, chari2003bluebox, shu2017long}
or a mixture of public and private datasets~\cite{maggi2010detecting, creech2014semantic}.

Prior critiques of network traces
are equally applicable to host traces.
For example,
several papers
bemoan how a lack of documentation prevents
replicable generation of traces~\cite{nehinbe2011critical, tavallaee2010toward, ringberg2008need}.
Deelman \etal~\cite{deelman2019initial} discuss how best-practices in cybersecurity, \eg
applying patches to address vulnerabilities, can change system functionality in ways that
might affect replicability, \eg by changing the code paths in the kernel that execute.
From their discussion, we can conclude that 
a replicable experiment must record not only the software that was used, 
but also the set of patches and updates that were applied to that software.
\xanthus~neatly sidesteps these problems by implicitly recording them by packaging the
entire environment into a virtual machine.

Other practical frameworks~\cite{jimenez2017popper} exist,
but these systems focus on re-running a computation to produce an identical output.
\xanthus' power lies in replicating a computation (\ie the training and test workloads)
specifically \emph{not} to produce an identical output, but to produce a different trace of
the same computation.
To the best of our knowledge,
\xanthus~is the first general framework that enables replication of workloads that interact
in complex ways with host operating systems. While we focus here on its use for evaluating
IDSes, \xanthus~can also be used to replicate results from experimental computer systems.

%% file: discussion.tex
Our \xanthus~prototype
is fully functional,
and we have already used it to evaluate \unicorn.
For example,
we used \xanthus~to
generate an APT trace (\autoref{sec:motivation:404:darpa}).
The traced APT attack
exploited a \texttt{wget} vulnerability (CVE-2016-4971)
to install a corrupt Debian package;
once installed,
the package contacted a command-and-control server,
and slowly exfiltrated data.
We were able to evaluate \unicorn's operation with three different provenance collection
infrastructures by changing only a few lines of code in the configuration script.
\xanthus~achieved the desired goals
from Section~\ref{sec:motivation:desiderata}:
  \begin{smitemize}
    \item \textbf{Replicability:} \xanthus~archives
          all of the information needed
          to recreate a trace.
          For example,
          \xanthus~records the environmental scripts
          provided by the user,
          and the contents of the corrupt Debian package.
          The Vagrant boxes that \xanthus~outputs
          are sufficient for a third party to
          replicate the original tracing conditions. 
    \item \textbf{Flexibility:} In \xanthus,
          the selection of an audit framework
          is orthogonal to
          the selection of the environmental conditions
          that drive the system behavior in the trace.
          This flexibility
          makes it easy to generate
          multiple datasets that use different logging mechanisms
          to observe the same environmental setup. This is the feature that allowed us to use
	  different provenance systems in our evaluation.
    \item \textbf{Longevity}: Currently, VMs are considered best practice for long-term
    	  digital preservation as the only requirement for running them is a compatible
	  hypervisor.
	  \xanthus~captures all necessary information inside a VM.
    \item \textbf{Usability:} \xanthus' script-based interface
          encourages the design of incremental, modular experiments.
          For example,
          as shown in~\autoref{sec:framework:setup},
          \xanthus~scripts enable a user to
          directly configure a VM and its applications.
          \xanthus~also directly integrates with
          popular penetration testing tools such as Metasploit
	  (\autoref{sec:framework:experimentation}),
          allowing off-the-shelf attacks
          to easily be added to a trace.
          \xanthus~re-runs an experiment
          using a single command.
    \item \textbf{Shareability:} \xanthus~automatically pushes
          VM images to VagrantCloud
          and the rest of a dataset archive to GitHub.
  \end{smitemize}
The interested reader
can find our APT dataset at \url{https://github.com/tfjmp/xanthus}.

\xanthus~improves dataset replicability,
but does not automatically improve dataset \textit{realism}.
\xanthus~users are responsible for ensuring
that environmental scripts and VM configurations
reflect plausible real-life scenarios.
Prior work on dataset fidelity~\cite{mchugh2000testing, sharafaldin2018toward}
can help \xanthus~users to create high-quality traces.

%% file: conclusion.tex
\xanthus~is a practical tool
for generating and sharing provenance traces.
By automating the minutiae of
trace collection and bundling,
\xanthus~enables the replicable evaluation of
host-based intrusion detectors.

%% file: availability.tex
\xanthus~is an open-source project,
available at \url{https://github.com/tfjmp/xanthus}
under an MIT license.
\xanthus's Ruby \texttt{gem} is freely distributed
at \url{https://rubygems.org/gems/xanthus}.
The CamFlow provenance capture system
is available at \url{http://camflow.org}
(GPL v2 license).
The SPADE provenance capture system
is available at \url{https://github.com/ashish-gehani/SPADE}
(GPL v3 license).

%% file: acknowledgements.tex
We acknowledge the support of the Natural Sciences and Engineering Research Council of Canada (NSERC).
Cette recherche a \'et\'e financ\'ee par le Conseil de recherches en sciences naturelles et en g\'enie du Canada (CRSNG).
This material is based upon work supported by the National Science Foundation under Grants ACI-1440800, ACI-1450277 and ACI-1547467. 
Any opinions, findings, and conclusions or recommendations expressed in this material are those of the authors and do not necessarily reflect the views of the National Science Foundation.

%% file: paper.bbl
\begin{thebibliography}{10}

\bibitem{darpatc}
{Transparent computing engagement 3 data release}, accessed \today.
\newblock \url{https://github.com/darpa-i2o/Transparent-Computing}.

\bibitem{unm}
{University of New Mexico system call dataset}, accessed \today.
\newblock \url{https://www.cs.unm.edu/~immsec/systemcalls.html}.

\bibitem{vagrantcloud}
{VagrantCloud}, accessed \today.
\newblock \url{https://app.vagrantup.com/boxes/search}.

\bibitem{balakrishnan2013opus}
Nikilesh Balakrishnan, Thomas Bytheway, Ripduman Sohan, and Andy Hopper.
\newblock Opus: A lightweight system for observational provenance in user
  space.
\newblock In {\em Workshop on the Theory and Practice of Provenance}. USENIX,
  2013.

\bibitem{bates2015trustworthy}
Adam Bates, Dave~Jing Tian, Kevin~RB Butler, and Thomas Moyer.
\newblock Trustworthy whole-system provenance for the linux kernel.
\newblock In {\em Security Symposium}, pages 319--334. USENIX, 2015.

\bibitem{chari2003bluebox}
Suresh~N Chari and Pau-Chen Cheng.
\newblock Bluebox: A policy-driven, host-based intrusion detection system.
\newblock {\em Transactions on Information and System Security}, 6(2):173--200,
  2003.

\bibitem{creech2013generation}
Gideon Creech and Jiankun Hu.
\newblock Generation of a new ids test dataset: Time to retire the kdd
  collection.
\newblock In {\em Wireless Communications and Networking Conference (WCNC)},
  pages 4487--4492. IEEE, 2013.

\bibitem{creech2014semantic}
Gideon Creech and Jiankun Hu.
\newblock A semantic approach to host-based intrusion detection systems using
  contiguous and discontiguous system call patterns.
\newblock {\em Transactions on Computers}, 63(4):807--819, 2014.

\bibitem{deelman2019initial}
Ewa Deelman, Victoria Stodden, Michela Taufer, and Von Welch.
\newblock Initial thoughts on cybersecurity and reproducibility.
\newblock In {\em International Workshop on Practical Reproducible Evaluation
  of Computer Systems}, pages 13--15, 2019.

\bibitem{elshafie2019improving}
Hussein~M Elshafie, Tarek~M Mahmoud, and Abdelmgeid~A Ali.
\newblock Improving the performance of the snort intrusion detection using
  clonal selection.
\newblock In {\em International Conference on Innovative Trends in Computer
  Engineering (ITCE)}, pages 104--110. IEEE, 2019.

\bibitem{gehani2012spade}
Ashish Gehani and Dawood Tariq.
\newblock Spade: Support for provenance auditing in distributed environments.
\newblock In {\em International Middleware Conference}, pages 101--120.
  ACM/IFIP/USENIX, 2012.

\bibitem{gregg2011dtrace}
Brendan Gregg and Jim Mauro.
\newblock {\em DTrace: Dynamic tracing in Oracle Solaris, Mac OS X, and
  FreeBSD}.
\newblock Prentice Hall Professional, 2011.

\bibitem{guo2012burrito}
Philip~J Guo and Margo Seltzer.
\newblock Burrito: Wrapping your lab notebook in computational infrastructure.
\newblock In {\em Workshop on the Theory and Practice of Provenance}. USENIX,
  2012.

\bibitem{haider2016windows}
Waqas Haider, Gideon Creech, Yi~Xie, and Jiankun Hu.
\newblock Windows based data sets for evaluation of robustness of host based
  intrusion detection systems (ids) to zero-day and stealth attacks.
\newblock {\em Future Internet}, 8(3):29, 2016.

\bibitem{haider2017generating}
Waqas Haider, Jiankun Hu, Jill Slay, Benjamin~P Turnbull, and Yi~Xie.
\newblock Generating realistic intrusion detection system dataset based on
  fuzzy qualitative modeling.
\newblock {\em Journal of Network and Computer Applications}, 87:185--192,
  2017.

\bibitem{han2020unicorn}
Xueyuan Han, Thomas Pasquier, Adam Bates, James Mickens, and Margo Seltzer.
\newblock Unicorn: Runtime provenance-based detector for advanced persistent
  threats.
\newblock In {\em Symposium on Network and Distributed System Security (NDSS)},
  2020.

\bibitem{han2017frappuccino}
Xueyuan Han, Thomas Pasquier, Tanvi Ranjan, Mark Goldstein, and Margo Seltzer.
\newblock Frappuccino: Fault-detection through runtime analysis of provenance.
\newblock In {\em Workshop on Hot Topics in Cloud Computing (HotCloud)}.
  USENIX, 2017.

\bibitem{han2018provenance}
Xueyuan Han, Thomas Pasquier, and Margo Seltzer.
\newblock Provenance-based intrusion detection: opportunities and challenges.
\newblock In {\em Workshop on the Theory and Practice of Provenance}. USENIX,
  2018.

\bibitem{hassan2019nodoze}
Wajih~Ul Hassan, Shengjian Guo, Ding Li, Zhengzhang Chen, Kangkook Jee, Zhichun
  Li, and Adam Bates.
\newblock Nodoze: Combatting threat alert fatigue with automated provenance
  triage.
\newblock In {\em Symposium on Network and Distributed System Security (NDSS)},
  2019.

\bibitem{hassan2018towards}
Wajih~Ul Hassan, Mark Lemay, Nuraini Aguse, Adam Bates, and Thomas Moyer.
\newblock Towards scalable cluster auditing through grammatical inference over
  provenance graphs.
\newblock In {\em Symposium on Network and Distributed System Security (NDSS)},
  2018.

\bibitem{illy2019securing}
Poulmanogo Illy, Georges Kaddoum, Christian~Miranda Moreira, Kuljeet Kaur, and
  Sahil Garg.
\newblock Securing fog-to-things environment using intrusion detection system
  based on ensemble learning.
\newblock In {\em Wireless Communications and Networking Conference (WCNC)},
  pages 1--7. IEEE, 2019.

\bibitem{jiang2006provenance}
Xuxian Jiang, AAron Walters, Dongyan Xu, Eugene~H Spafford, Florian Buchholz,
  and Yi-Min Wang.
\newblock Provenance-aware tracing of worm break-in and contaminations: A
  process coloring approach.
\newblock In {\em International Conference on Distributed Computing Systems
  (ICDCS)}, pages 38--38. IEEE, 2006.

\bibitem{jiang2005virtual}
Xuxian Jiang, Dongyan Xu, Helen~J Wang, and Eugene~H Spafford.
\newblock Virtual playgrounds for worm behavior investigation.
\newblock In {\em International Workshop on Recent Advances in Intrusion
  Detection}, pages 1--21. Springer, 2005.

\bibitem{jimenez2017popper}
Ivo Jimenez, Michael Sevilla, Noah Watkins, Carlos Maltzahn, Jay Lofstead,
  Kathryn Mohror, Andrea Arpaci-Dusseau, and Remzi Arpaci-Dusseau.
\newblock The popper convention: Making reproducible systems evaluation
  practical.
\newblock In {\em Parallel and Distributed Processing Symposium Workshops
  (IPDPSW)}, pages 1561--1570. IEEE, 2017.

\bibitem{kennedy2011metasploit}
David Kennedy, Jim O'gorman, Devon Kearns, and Mati Aharoni.
\newblock {\em Metasploit: The penetration tester's guide}.
\newblock No Starch Press, 2011.

\bibitem{king2007introduction}
Gary King.
\newblock An introduction to the dataverse network as an infrastructure for
  data sharing, 2007.

\bibitem{lichodzijewski2002host}
Peter Lichodzijewski, A~Nur Zincir-Heywood, and Malcolm~I Heywood.
\newblock Host-based intrusion detection using self-organizing maps.
\newblock In {\em International Joint Conference on Neural Networks}, volume~2,
  pages 1714--1719. IEEE, 2002.

\bibitem{lippmann20001999}
Richard Lippmann, Joshua~W Haines, David~J Fried, Jonathan Korba, and Kumar
  Das.
\newblock The 1999 darpa off-line intrusion detection evaluation.
\newblock {\em Computer Networks}, 34(4):579--595, 2000.

\bibitem{liu2018towards}
Yushan Liu, Mu~Zhang, Ding Li, Kangkook Jee, Zhichun Li, Zhenyu Wu, Junghwan
  Rhee, and Prateek Mittal.
\newblock Towards a timely causality analysis for enterprise security.
\newblock In {\em Symposium on Network and Distributed System Security (NDSS)},
  2018.

\bibitem{maggi2010detecting}
Federico Maggi, Matteo Matteucci, and Stefano Zanero.
\newblock Detecting intrusions through system call sequence and argument
  analysis.
\newblock {\em Transactions on Dependable and Secure Computing}, 7(4):381--395,
  2010.

\bibitem{mahoney2003analysis}
Matthew~V Mahoney and Philip~K Chan.
\newblock An analysis of the 1999 darpa/lincoln laboratory evaluation data for
  network anomaly detection.
\newblock In {\em International Workshop on Recent Advances in Intrusion
  Detection}, pages 220--237. Springer, 2003.

\bibitem{manzoor2016fast}
Emaad Manzoor, Sadegh~M Milajerdi, and Leman Akoglu.
\newblock Fast memory-efficient anomaly detection in streaming heterogeneous
  graphs.
\newblock In {\em SIGKDD International Conference on Knowledge Discovery and
  Data Mining}, pages 1035--1044. ACM, 2016.

\bibitem{mchugh2000testing}
John McHugh.
\newblock Testing intrusion detection systems: a critique of the 1998 and 1999
  darpa intrusion detection system evaluations as performed by lincoln
  laboratory.
\newblock {\em ACM Transactions on Information and System Security},
  3(4):262--294, 2000.

\bibitem{milajerdi2019holmes}
Sadegh~M Milajerdi, Rigel Gjomemo, Birhanu Eshete, R~Sekar, and
  VN~Venkatakrishnan.
\newblock Holmes: real-time apt detection through correlation of suspicious
  information flows.
\newblock In {\em Symposium on Security and Privacy (SP)}, pages 1137--1152.
  IEEE, 2019.

\bibitem{moore2012metasploitable}
HD~Moore.
\newblock Metasploitable 2 exploitability guide.
\newblock {\em Retrieved June}, 27:2013, 2012.

\bibitem{muniswamy2006provenance}
Kiran-Kumar Muniswamy-Reddy, David~A Holland, Uri Braun, and Margo~I Seltzer.
\newblock Provenance-aware storage systems.
\newblock In {\em Annual Technical Conference}, pages 43--56. USENIX, 2006.

\bibitem{murtaza2013host}
Syed~Shariyar Murtaza, Wael Khreich, Abdelwahab Hamou-Lhadj, and Mario Couture.
\newblock A host-based anomaly detection approach by representing system calls
  as states of kernel modules.
\newblock In {\em International Symposium on Software Reliability Engineering
  (ISSRE)}, pages 431--440. IEEE, 2013.

\bibitem{national2019reproducibility}
{National Academies of Sciences, Engineering, and Medicine} et~al.
\newblock {\em Reproducibility and replicability in science}.
\newblock National Academies Press, 2019.

\bibitem{nehinbe2011critical}
Joshua~Ojo Nehinbe.
\newblock A critical evaluation of datasets for investigating idss and ipss
  researches.
\newblock In {\em International Conference on Cybernetic Intelligent Systems
  (CIS)}, pages 92--97. IEEE, 2011.

\bibitem{pasquier2017practical}
Thomas Pasquier, Xueyuan Han, Mark Goldstein, Thomas Moyer, David Eyers, Margo
  Seltzer, and Jean Bacon.
\newblock Practical whole-system provenance capture.
\newblock In {\em Symposium on Cloud Computing}, pages 405--418. ACM, 2017.

\bibitem{pasquier2018ccs}
Thomas Pasquier, Xueyuan Han, Thomas Moyer, Adam Bates, Olivier Hermant, David
  Eyers, Jean Bacon, and Margo Seltzer.
\newblock Runtime analysis of whole-system provenance.
\newblock In {\em Conference on Computer and Communications Security (CCS)}.
  ACM, 2018.

\bibitem{pendleton2017dataset}
Marcus Pendleton and Shouhuai Xu.
\newblock A dataset generator for next generation system call host intrusion
  detection systems.
\newblock In {\em Military Communications Conference (MILCOM)}, pages 231--236.
  IEEE, 2017.

\bibitem{pohly2012hi}
Devin~J Pohly, Stephen McLaughlin, Patrick McDaniel, and Kevin Butler.
\newblock {Hi-Fi: Collecting high-fidelity whole-system provenance}.
\newblock In {\em Annual Computer Security Applications Conference}, pages
  259--268. ACM, 2012.

\bibitem{ringberg2008need}
Haakon Ringberg, Matthew Roughan, and Jennifer Rexford.
\newblock The need for simulation in evaluating anomaly detectors.
\newblock {\em SIGCOMM Computer Communication Review}, 38(1):55--59, 2008.

\bibitem{sharafaldin2018toward}
Iman Sharafaldin, Arash~Habibi Lashkari, and Ali~A Ghorbani.
\newblock Toward generating a new intrusion detection dataset and intrusion
  traffic characterization.
\newblock In {\em International Conference on Information Systems Security and
  Privacy (ICISSP)}, pages 108--116, 2018.

\bibitem{shiravi2012toward}
Ali Shiravi, Hadi Shiravi, Mahbod Tavallaee, and Ali~A Ghorbani.
\newblock Toward developing a systematic approach to generate benchmark
  datasets for intrusion detection.
\newblock {\em Computers \& Security}, 31(3):357--374, 2012.

\bibitem{shu2017long}
Xiaokui Shu, Danfeng~Daphne Yao, Naren Ramakrishnan, and Trent Jaeger.
\newblock Long-span program behavior modeling and attack detection.
\newblock {\em Transactions on Privacy and Security (TOPS)}, 20(4):12, 2017.

\bibitem{spillane2009story}
Richard~P Spillane, Russell Sears, Chaitanya Yalamanchili, Sachin Gaikwad,
  Manjunath Chinni, and Erez Zadok.
\newblock Story book: An efficient extensible provenance framework.
\newblock In {\em Workshop on the Theory and Practice of Provenance}. USENIX,
  2009.

\bibitem{tavallaee2010toward}
Mahbod Tavallaee, Natalia Stakhanova, and Ali~Akbar Ghorbani.
\newblock Toward credible evaluation of anomaly-based intrusion-detection
  methods.
\newblock {\em Transactions on Systems, Man, and Cybernetics}, 40(5):516--524,
  2010.

\bibitem{wang2020you}
Qi~Wang, Wajih~Ul Hassan, Ding Li, Kangkook Jee, Xiao Yu, Kexuan Zou, Junghwan
  Rhee, Zhengzhang Chen, Wei Cheng, C~Gunter, et~al.
\newblock You are what you do: Hunting stealthy malware via data provenance
  analysis.
\newblock In {\em Symposium on Network and Distributed System Security (NDSS)},
  2020.

\end{thebibliography}
